\documentclass[twocolumn,showpacs,preprintnumbers,amsmath,amssymb]{revtex4}

\usepackage{graphicx}
\usepackage{dcolumn,exscale}
\usepackage{bm}

\newcommand{\opI}{\hat{I}}

\newcommand{\opS}{\hat{S}}

\newcommand{\opE}{\hat{1}}

\newcommand{\opU}{\hat{U}}

\newcommand{\oprho}{\hat{\rho}}

\begin{document}

\preprint{malonlett}

\title{Entanglement between an electron and a nuclear spin $\mathbf{\frac{1}{2}}$}

\author{M.~Mehring}
\author{J.~Mende}
\author{W.~Scherer}
\affiliation{2.~Physikalisches~Institut, Universit\"at~Stuttgart,\\
  Pfaffenwaldring~57, 70550 Stuttgart, Germany}

\date{\today}

\begin{abstract}
We report on the preparation and detection of entangled states between an electron spin 1/2 and a nuclear spin 1/2 in a
molecular single crystal. These were created by applying pulses at ESR (9.5 GHz) and NMR (28 MHz)
frequencies. Entanglement was detected by using a special entanglement detector sequence based on a unitary back
transformation including phase rotation.
\end{abstract}

\maketitle

The entanglement between two spins 1/2 is at the heart of quantum mechanics. Ever since
 a so-called "paradox" was formulated by Einstein, Podolsky and Rosen (EPR) \cite{einstein:35}, referring
to local measurements performed on the individual spins of a delocalized entangled pair, properties
 of entanglement and its consequences for quantum physics has been discussed in great
 detail \cite{greenberger:89}. In the context of quantum information processing (QIP) entanglement
 has been considered as a resource for quantum parallelism (speedup of quantum computing)
 \cite{deutsch:85,deutsch:92,grover:97} and quantum cryptography \cite{bennet:84,bennet:92}. A number of
 these quantum algorithms have been demonstrated in NMR (nuclear magnetic resonance)
 quantum computing \cite{cory97, gershenfeld:97, knill:98}.

\begin{figure}[htb]
            \centerline{\includegraphics[width=0.4\textwidth]{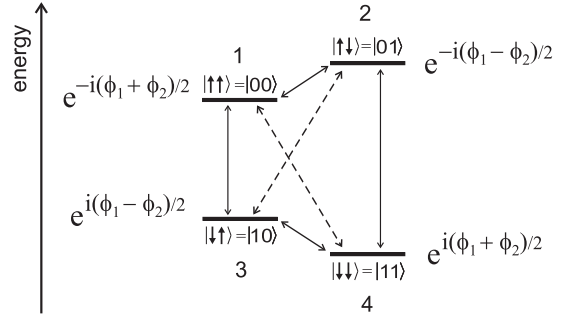}}
            \caption{Schematic diagram of the four energy levels of a two spin system with $S=1/2$ and $I=1/2$. The
            solid arrows denote allowed transitions. The dotted arrows indicate forbidden transitions,
            corresponding to entangled states. The phase dependence of the quantum states under
            $z$-rotations is also indicated (see text).\label{4niveau}}
    \end{figure}

In this contribution we report on the experimental preparation and observation of the
entangled states of an electron spin $S = 1/2$ and a nuclear spin $I = 1/2$
in a crystalline solid. The spins considered here are a proton and a radical (unpaired electron spin) produced
 by x-ray irradiation of a malonic acid single crystal \cite{mcconnell:60}. This leads to the partial conversion of
  the CH$_2$ group of the malonic acid molecule to the radical $^{\bullet}$CH where the dot marks
 the electron spin. In a strong magnetic field the following four Zeeman product states
 $|m_Sm_I\rangle = ~|\uparrow\uparrow\rangle,~|\uparrow\downarrow\rangle,~|\downarrow\uparrow\rangle,
 ~|\downarrow\downarrow\rangle$ exist where the arrows label the $\pm 1/2$ states of the electron
  and the nuclear spin. Equivalently we will use a qubit labelling as $|m_Sm_I\rangle =
  |00\rangle,~|01\rangle,~|10\rangle,~|11\rangle$.
 The energy level diagram corresponding to the electron-proton spin system is shown in fig. \ref{4niveau}, where
 we have also indicated the possible
  ESR ($\Delta m_S =\pm1$) and NMR transitions ($\Delta m_I =\pm1$) of the individual spins by solid arrows.

What we are aiming at are states of
the type
    \begin{equation}\label{Bell}
      \Psi^\pm = \frac{1}{\sqrt 2}\left(\mid \uparrow\downarrow\rangle\pm \mid
      \downarrow\uparrow\rangle\right)
      \mbox{ and } \Phi^\pm = \frac{1}{\sqrt 2}\left(\mid \uparrow\uparrow\rangle\pm \mid
      \downarrow\downarrow\rangle\right).
    \end{equation}
These represent all four possible entangled states of a two qubit system, also called the Bell states
 of two spins 1/2. They correspond to a superposition of the states in fig.~\ref{4niveau}
 connected by dashed arrows.

Electron spin resonance (ESR) was performed at X-band (9.49 GHz) at $T = 40$~K. The low temperature was
chosen only for reasons of signal-to-noise ratio. The two well resolved ESR lines
due to the $^\bullet$CH proton depend on the orientation of the single crystal and were observed at magnetic fields of
338.2~mT and 339.2~mT with a linewidth of 0.5~mT for the ESR and about 1~MHz for the ENDOR
(electron nuclear double resonance) line.
This orientation corresponds to a principal axis of the hyperfine tensor. There are two different
 proton NMR transitions. We applied pulsed ENDOR techniques to one of them at the
frequency of 28.05~MHz.

In the high temperature approximation we express the Boltzmann spin density matrix as
 $\oprho_\mathrm{B} = (1-K_\mathrm{B})\frac{1}{4}\opE+K_\mathrm{B}\cdot\oprho_\mathrm{P}$ with
 $K_\mathrm{B} =\mu_\mathrm{B}B_0/k_\mathrm{B}T$ (for $g = 2$)
 and where the pseudo Boltzmann density matrix is defined as $\oprho_\mathrm{P}=(\frac{1}{4}\opE-\frac{1}{2}\opS_z)$
  which corresponds to an equal polulation of the states $|\downarrow\uparrow\rangle$ and
  $|\downarrow\downarrow\rangle$ ($p_3=p_4=1/2$) and equivalently $|\uparrow\downarrow\rangle$ and
  $|\uparrow\uparrow\rangle$ ($p_1=p_2=0$) with $\mathrm{tr}\{\oprho_\mathrm{P}\}=1$.
We used here the assumption that the Larmor frequency of the nuclear spin $\omega_{0I}$ is much smaller than
 the Larmor frequency of the electron spin $\omega_{0S}$. Note that the pseudo-pure density matrix $\oprho_{00}$
 can be expressed as $\oprho_{00} = \frac{1}{4}\opE+\frac{1}{2}\opS_z+\frac{1}{2}\opI_z+\opS_z\opI_z=|00\rangle\langle 00|$
corresponding to the pure state $\psi = |00\rangle$.

In what follows we will prepare density matrices corresponding to the Bell states according to eqn. (\ref{Bell}).
Since we will apply selective transitions we need to consider only a three level subsystem for each of the Bell states.
In order to prepare all four states $\Psi^\pm$ and $\Phi^\pm$ we only need to apply transition selective excitations
to either of the three level subsystems 1,2,3; 1,2,4; 1,3,4 or 2,3,4. Here we restrict ourselves to the sublevels 1,2,4 for
 creating the $\Phi^\pm$ states and 1,2,3 to create the $\Psi^\pm$ states. As an example we treat
 the $\Psi^-$ state in detail. It corresponds to the well known Einstein Podolsky Rosen (EPR) state \cite{einstein:35}.
The preparation of $\Psi^-$ proceeds by first applying a selective $\pi$-pulse to the $1\leftrightarrow3$ transition
to create the following pseudo pure populations:  $p_1= 1/2$, $p_2= 0$, $p_3= 0$ of the 1,2,3 three level subsystem.
 The corresponding pseudo pure density matrix of the three-level subsystem (1,2,3) represents the pseudo pure
 state $\mid\uparrow\uparrow\rangle$. The creation of the $\Psi^{\pm}$ states can now be achieved
 by the pulse sequence $P^{12}_I(\mp\pi/2)$ followed by $P^{13}_S(-\pi)$ where the plus sign in $P^{12}_I(\mp\pi/2)$
 creates the $\Psi^-$ state. Here we use the abbreviation $P^{jk}_{S,I}(\beta)$ for a selective pulse at the transition
  $j\leftrightarrow k$ with rotation angle $\beta$. The label $S$ refers to an electrons spin transition,
  whereas the label $I$ stands for a nuclear spin transition. The corresponding unitary transformation results in

\begin{equation}\label{psi}
|\uparrow\uparrow\rangle \stackrel{P^{12}_I(\pi/2)}{\longrightarrow}\frac{1}{\sqrt{2}}(|\uparrow\uparrow\rangle
+|\uparrow\downarrow\rangle) \stackrel{P^{13}_S(-\pi)}{\longrightarrow}\frac{1}{\sqrt{2}}(|\uparrow\downarrow\rangle
-|\downarrow\uparrow\rangle).
\end{equation}

In order to create $\Psi^+$ the plus sign in $P^{13}_I(\pm\pi)$ must be chosen. For completeness we note that
the $\Phi^{\pm}$ states can be created by following the same line of reasoning when starting from the sublevels 1,2,4
with a preparatory $\pi$-pulse at the $2\leftrightarrow4$ transition. Next the pulse sequence $P^{12}_I(\pm\pi/2)$
followed by $P^{24}_S(-\pi)$ is applied creating the $\Phi^{\pm}$ state except an overall minus sign.
Other scenarios using the other sublevels are possible and will be presented in a more extensive publication.

In order to prove that the $\Psi^-$ state has indeed been created we apply a density matrix tomography which is based on
the phase dependence of the entangled state as already sketched in fig.~\ref{4niveau}. The phase factors noted
there represent the phase dependence of the corresponding states under rotation about the quantization axis (z-axis).
This corresponds to the unitary transformations $\opU_{S_z}=\exp(-i\phi_1\opS_z)$ and $\opU_{I_z}=\exp(-i\phi_2\opI_z)$
leading to $\opU_{S_z}\opU_{I_z}|m_Sm_I\rangle = \exp(-i(\phi_1m_S+\phi_2m_I))|m_Sm_I\rangle$.
A single ESR transition ($\Delta m_S = \pm 1$) will have a phase dependence $\phi_1$ under z-axis rotation, whereas a single
 NMR transition ($\Delta m_I = \pm 1$) will have a phase dependence $\phi_2$. Each of the entangled states
 $\Psi^\pm$ and $\Phi^\pm$ is characterized
 by a linear combination $\phi_1 \pm \phi_2$ of both phases. This is another manifestation of the fact, that these states
  are global states and no local measurement on the single qubits reveals any information about the entangled state.

Since the entangled state is not directly observable we need to transform it to an observable state.
Our entangled state detector therefore corresponds to a unitary back transformation comprised of e.g. $P^{13}_S(-\pi)$
followed by $P^{12}_I(-\pi/2)$ applied to the state $\Psi^-$. In order to distinguish entangled states from other superposition states we encode
their characters in the phase dependence under z-rotation as discussed before. Therefore we apply phase shifted pulses
 of the type $P^{13}_S(-\pi,~\phi_1)$ and $P^{12}_I(-\pi/2,~\phi_2)$ which corresponds to a rotation about the z-axis by
  the angles $\phi_1$ and $\phi_2$. The complete $\Psi^{\pm}$ detector
 sequence now reads
\begin{equation}\label{UdetPsi}
\opU^\Psi_\mathrm{d}(\phi_1,~\phi_2)= P^{12}_I(-\pi/2,~\phi_2)\,P^{13}_S(-\pi,~\phi_1).
\end{equation}
Since our observable is the ESR transition intensity, detected via an electron spin echo, the entangled
state tomography corresponds to the evaluation of the following signal strength for $\Psi^-$

\begin{equation}\label{entangdetPsi}
S^\Psi_\mathrm{d}(\phi_1,\phi_2) = \mathrm{tr}\left\{2\opS^{13}_z\,\opU^\Psi_\mathrm{d}\,|\Psi^-\rangle\langle\Psi^-|\,\opU^{\Psi\dag}_\mathrm{d}\right\}
\end{equation}

where $\opS^{13}_z$ is the fictitious spin 1/2 of the $1\leftrightarrow 3$ transition.
With the current definitions we obtain the expression

\begin{equation}\label{detPsi}
S^\Psi_\mathrm{d}(\phi_1,\phi_2) = \frac{1}{2}\big(1-\cos(\phi_1-\phi_2)\big).
\end{equation}

Alternatively we have used for the detection of the $\Phi^{\pm}$ states the sequence
\begin{equation}\label{UdetPhi}
\opU^\Phi_\mathrm{d}(\phi_1,~\phi_2)= P^{12}_I(-\pi/2,~\phi_2)\,P^{24}_S(-\pi,~\phi_1)
\end{equation}

leading to a detector signal for $\Phi^+$

\begin{equation}\label{entangdetPhi}
S^\Phi_\mathrm{d}(\phi_1,\phi_2) = \mathrm{tr}\left\{2\opS^{24}_z\,\opU^\Phi_\mathrm{d}\,|\Phi^+\rangle\langle\Phi^+|\,\opU^{\Phi\dag}_\mathrm{d}\right\}.
\end{equation}

with

\begin{equation}\label{detPhi}
S^\Phi_\mathrm{d}(\phi_1,\phi_2) = \frac{1}{2}\big(1-\cos(\phi_1+\phi_2)\big).
\end{equation}

A more detailed discussion of the phase dependence of the detector signal will be presented in a more extended publication.

\begin{figure}
  \includegraphics [width=0.3\textwidth]{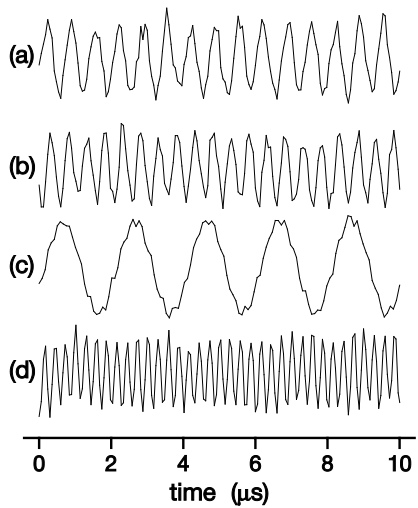}
  \caption{Phase interferograms versus time $n~\Delta t$ for four different sets of experiments. (a) $\phi_1=0$, (b) $\phi_2=0$,
   (c) $\phi_1\neq 0,~\phi_2\neq 0$ for the entangled state $ \Psi^-$ with phase dependence $|\phi_1-\phi_2|$
   (see eqn. \ref{detPsi}).
   (d) $\phi_1\neq 0,~\phi_2\neq 0$ for entangled state $\Phi^+$ with phase dependence $|\phi_1+\phi_2|$
   (see eqn. \ref{detPhi})
     \label{vszphase}}
\end{figure}

The phase shifts were implemented by incrementing the phase of the
individual detection pulses in consecutive experiments according to $\phi_j(n) = \Delta\omega_j\,n\,\Delta t$ with $j=1,2$. The artificial
phase frequencies $\Delta\omega_j= 2\pi\Delta\nu_j$ were arbitrary chosen as $\nu_1=2.0$~MHz and $\nu_2=1.5$~MHz. Examples of different phase increments
are shown in fig.~\ref{vszphase}.

\begin{figure}
  \includegraphics[width=0.3\textwidth]{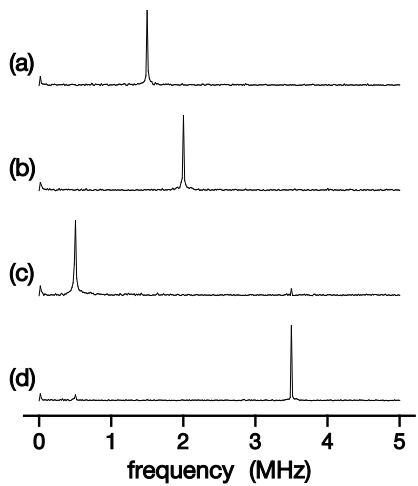}
  \caption{Fourier transform of the phase interferograms shown in fig.~\ref{vszphase} for phase frequencies $\nu_1=2.0$~MHz
   and $\nu_2=1.5$~MHz (see text). (a) $\nu_2=1.5$~MHz,
  (b) $\nu_1=2.0$~MHz, (c) $|\nu_1-\nu_2|$,
  (d) $|\nu_1+\nu_2|$.
  \label{vszphasef}}
\end{figure}

Four different sets of phase variations were chosen to demonstrate the individual and combined phase frequencies.
In fig.~\ref{vszphase} a, b we have set (a) $\nu_1=0$ or (b) $\nu_2=0$ in order to demonstrate the $\phi_2$ and $\phi_1$
dependence as a reference. The corresponding spectra (see fig.~\ref{vszphasef} a, b) are obtained after Fourier transformation.
These would also be observed for non-entangled superposition states of either ESR ($\nu_1$) or NMR transitions ($\nu_2$).
The characteristics of the entangled states shows up in the combined phase dependence (see eqns. \ref{detPsi} and \ref{detPhi}). This is demonstrated
for the $\Psi^-$ state in fig.~\ref{vszphase} c where the interferogram already shows the phase difference behavior
which is even more clearly evident from the spectral representation of fig.~\ref{vszphasef}c showing a line at the
 difference frequency $\nu_1-\nu_2$. In a similar way the $\Phi^{\pm}$ states
 were created which under the corresponding tomography sequence show a phase dependence as $\phi_1+\phi_2$ which is
 evident from the interferogram (fig.~\ref{vszphase}d) and more clearly from the phase spectrum (fig.~\ref{vszphasef}d)
 leading to a spectral line at $\nu_1+\nu_2$.

Here we have used an ensemble of electron-nuclear spin pairs with a mixed state density matrix. Like in all so far
published NMR quantum computing experiments the corresponding entangled state would be better termed {\it pseudo
entangled states} \cite{cory97, gershenfeld:97, knill:98}. However, we point out that the same pulse sequences could be applied to a pure state
electron-nuclear spin pair in order to create the discussed entangled states. Also the same Bell state detection
sequences proposed here would apply. In fact the same phase dependence would be observed for pure states.
Moreover, the experiments presented here would reach the quantum limit (see Warren et al. \cite{warren:97}) according to the
PPT (positive partial transpose) criterion \cite{peres:96, horodecki:96} at $\tanh(\hbar\omega_S/2k_BT) =\sqrt{2}$
which corresponds to a temperature $T_Q$ =  2.576~K for an ESR frequency of 95~GHz.

Theoretically the scenario reported here might seem to be obvious, however, experimentally this type of experiment
 is rather demanding because of the extreme difference of the ESR and NMR linewidth and the duration of the microwave
 and radio-frequency pulses. We have used the following typical values: $32$~ns for the ESR and $1.6~\mu$s for
  the NMR $\pi$-pulses. Entanglement is achieved with these in $832$~ns. However, because of the broad ESR and NMR lines a complete excitation of the transitions
   could not be obtained with these pulses. This leads to errors in the effective rotation angles of the corresponding pulses. As a consequence
 of this an incomplete creation of the entangled states results, leading to residual separable states. These
 are expected to vary as $\phi_1$ and $\phi_2$. In order to estimate this effect we calculate the resulting
 phase dependence of the detector signal (electron spin echo) for the deviation $\delta_j$ of the $\pi/2$-pulses at
 ESR ($\delta_1$) and NMR ($\delta_2$) frequencies which results in

\begin{eqnarray}\label{entangdeterr}
S_\mathrm{d}(\phi_1,\,\phi_2) &=& a_0 + a_1\cos\phi_1+a_2\cos\phi_2 \nonumber \\
                              &&+a_{12}\cos(\phi_1-\phi_2)
\end{eqnarray}

with $a_1=-\frac{1}{2}\delta_1(1-\delta_1)\delta^2_2$, $a_2=-\frac{1}{4}\delta_1\delta_2$ and
$a_{12}=\frac{1}{4}(1-\delta_1^2)(1-\delta_2^2)$ up to fourth order.

We note that the pulse errors introduce $\cos\phi_j$ (j=1,2) dependencies in addition to the expected $\cos(\phi_1-\phi_2)$
 dependence of the entangled state. In order to reduce the $\cos\phi_j$ dependencies in the detector signal we
 have applied a phase cycling procedure where the two signals $S_d(\phi_1,\,\phi_2)$ and $S_d(\phi_1+\pi,\,\phi_2+\pi)$
 were added. This eliminates the $\cos\phi_j$ dependencies but retains the $\cos(\phi_1-\phi_2)$ dependence of the
 entangled state. This procedure was applied to the data presented in figs.~\ref{vszphase} and \ref{vszphasef}.

We acknowledge a financial support by the German Bundesministerium f\"ur Bildung und Forschung (BMBF).

\bibliographystyle{apsrev}

\end{document}